\documentclass{sigchi-ext}
\usepackage[T1]{fontenc}
\usepackage{textcomp}
\usepackage[scaled=.92]{helvet} 
\usepackage{graphicx} 
\usepackage{balance}  
\usepackage{booktabs} 
\usepackage{ccicons}  
\usepackage{ragged2e} 

\copyrightinfo{Permission to make digital or hard copies of all or
part of this work for personal or classroom use is granted without
fee provided that copies are not made or distributed for profit or
commercial advantage and that copies bear this notice and the full
citation on the first page. Copyrights for components of this work
owned by others than ACM must be honored. Abstracting with credit is
permitted. To copy otherwise, or republish, to post on servers or to
redistribute to lists, requires prior specific permission and/or a
fee. Request permissions from permissions@acm.org.\\
{\emph{UbiComp/ISWC'18 Adjunct}}, October 8--12, 2018, Singapore, Singapore. \\
Copyright \copyright~2018 ACM ISBN/14/04...\$15.00. \\
https://doi.org/10.1145/3267305.3267684}

\def\plaintitle{Ubiquitous Event Mining to Enhance Personal Health} 
\def\emptyauthor{}
\def\plainkeywords{Event Mining; Pattern Recognition, Personal Health Navigation, User Interface}

\title{Ubiquitous Event Mining to Enhance Personal Health}

\numberofauthors{3}
\author{%
  \alignauthor{%
    \textbf{Vaibhav Pandey}\\
    \affaddr{University of California, Irvine} \\
    \affaddr{Irvine, California, \newline United States of America} \\
    \email{vaibhap1@uci.edu} }
   \vfil \alignauthor{%
    \textbf{Nitish Nag}\\
    \affaddr{University of California, Irvine}\\
    \affaddr{Irvine, California, \newline United States of America}\\
    \email{nagn@uci.edu}}
    \vfil \alignauthor{%
    \textbf{Ramesh Jain}\\
    \affaddr{University of California, Irvine}\\
    \affaddr{Irvine, California, \newline United States of America}\\
    \email{jain@ics.uci.edu} } }

\definecolor{linkColor}{RGB}{6,125,233}
\hypersetup{%
  pdftitle={\plaintitle},
  pdfauthor={\emptyauthor},
  pdfkeywords={\plainkeywords},
  bookmarksnumbered,
  pdfstartview={FitH},
  colorlinks,
  citecolor=black,
  filecolor=black,
  linkcolor=black,
  urlcolor=linkColor,
  breaklinks=true,
}

\begin{document}

\CopyrightYear{2018} 
\setcopyright{acmcopyright} 
\conferenceinfo{UbiComp/ISWC'18 Adjunct,}{October 8--12, 2018, Singapore, Singapore}
\isbn{978-1-4503-5966-5/18/10}\acmPrice{$15.00}
\doi{https://doi.org/10.1145/3267305.3267684}


\maketitle

\RaggedRight{} 


\begin{abstract}
Advances in user interfaces, pattern recognition, and ubiquitous computing continue to pave the way for better navigation towards our health goals. Quantitative methods which can guide us towards our personal health goals will help us optimize our daily life actions, and environmental exposures. Ubiquitous computing is essential for monitoring these factors and actuating timely interventions in all relevant circumstances. We need to combine the events recognized by different ubiquitous systems and derive actionable causal relationships from an event ledger. Understanding of user habits and health should be teleported between applications rather than these systems working in silos, allowing systems to find the optimal guidance medium for required interventions. We propose a method through which applications and devices can enhance the user experience by leveraging event relationships, leading the way to more relevant, useful, and, most importantly, pleasurable health guidance experience. 
 
\end{abstract}




\section{Introduction}
Humans have always excelled at recognizing patterns in their surroundings and utilizing them in future decision making. These patterns are observed and enriched throughout the life of an individual and shared across the community, validating some patterns as significant (or causal) and rejecting some as coincidences. This collective knowledge pool is shared and further enriched across successive generations, allowing humankind to make progress in science and technology \cite{HarariSapiensHumankind}.


\begin{figure}
  \includegraphics[width=0.833\columnwidth]{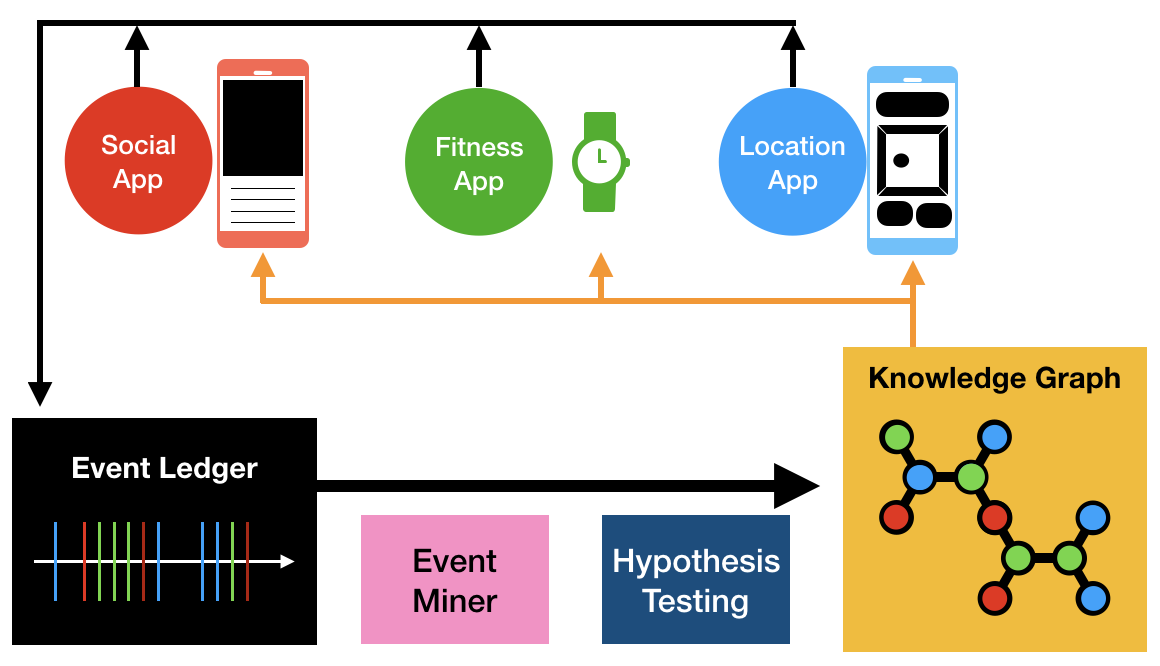}
  \caption{As devices and applications interact with a user, a timeline is created of all events pertinent to the user. Event mining establishes relationships on a temporal scale between events. These correlations provide the potential hypothesis that we may want to test on a user to produce a given outcome. After we are able to implement the hypothesis testing, we can start building a knowledge graph of events. This common set of personalized event knowledge can then be consumed by other applications or devices in real-time to continue to serve the user.}~\label{fig:flow}
\end{figure}

Similar approaches can be used for ubiquitous computing and designing effective user interventions through smartphones, wearables, and IoT devices. We are constantly monitoring our life using a variety of devices and applications allowing us to estimate the state of different aspects of our life, health, and habits \cite{Nag2018Cross-ModalEstimation,Oh2017FromChronicles}. We can take further advantage of this ubiquitous computing paradigm to develop a navigational approach to health, in which our computing services cater to our individual preferences and help actuate actions that lead us towards our health goals. 
\newline
Although ubiquitous systems have operated independently and effectively in their own silos, there is still a lack of communication and knowledge sharing between these systems. By pooling the events recognized by these systems, we can discover how different aspects of our life are interconnected and how they are influencing our health state. Mining relevant patterns from the event pool can find causal relationships between these events, physiological parameters, and behaviors. Such a framework would also allow us to derive rich insights about the individual and ultimately lead to effective lifestyle interventions across various mediums of computing intervention (fig. \ref{fig:flow}). Since health is a product of all of our daily life actions, the more our computational assistance can be personalized, the better our guidance towards our health goals. One method through which we can personalize this is by ensuring that all of our devices and applications understand our historical patterns as individuals. In order for this to be useful, we need event mining to elucidate these patterns.

Finding the associations between events is the first step towards building the personal knowledge graph that can be used by different applications. We will need to implement a "do-operator" to attribute the outcome to the associated cause \cite{Pearl2018TheEffect}. Once the causal relationship has been established, the event relationship can be added to the knowledge graph. Building the knowledge graph can be viewed as an exploratory step, in which we are exploring all the possible causes for an event. Similarly, guidance can be viewed as exploiting these relationships, in which different devices and applications can find a causal rule relevant in the given context and provide recommendations.

\section{Explore: Building the knowledge graph}
\begin{marginfigure}[1pc]
  \begin{minipage}{\marginparwidth}
    \centering
    \includegraphics[width=0.9\marginparwidth]{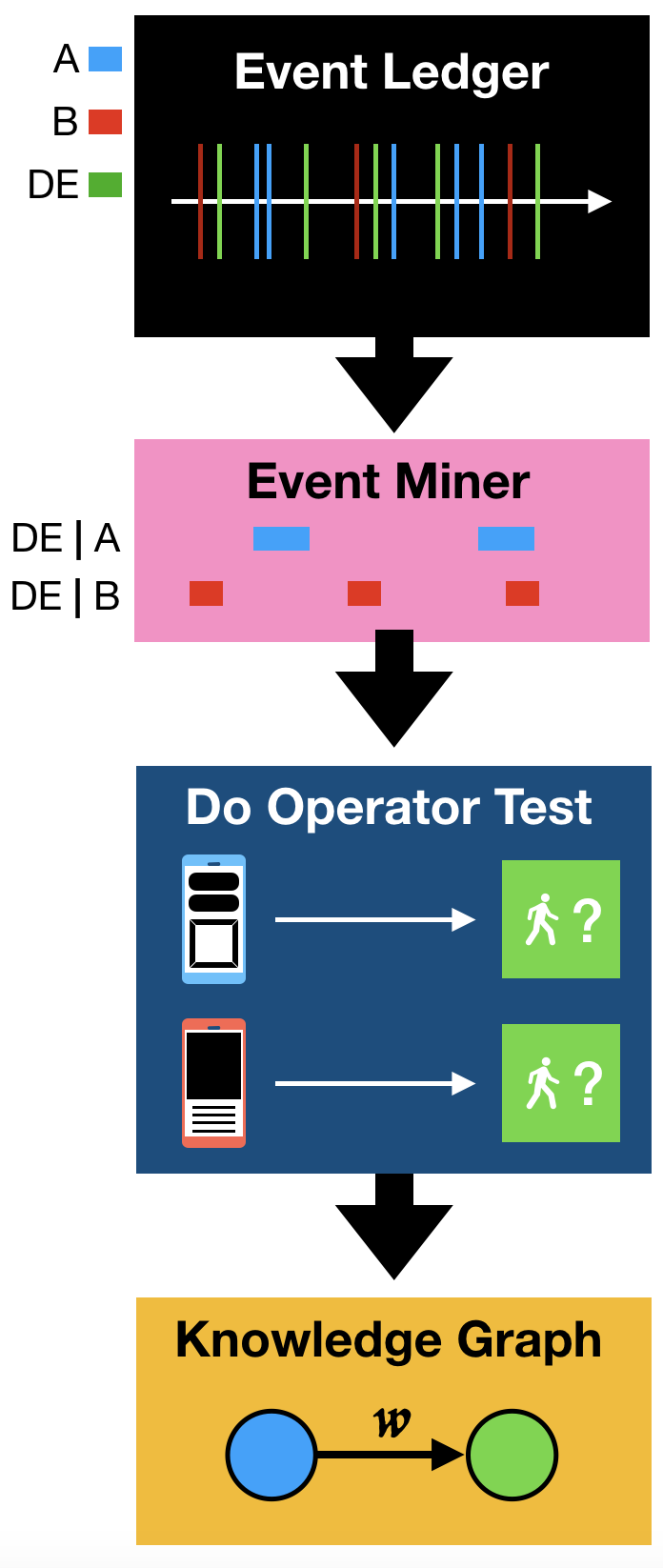}
    \caption{When we see relationships between Event A (A) or Event B (B) and a Desired Event (DE), we can test if there is a causal relationship by implementing an A/B Testing for the user. If the user responds positively towards a certain test, we can add a confidence weighted edge between two events in the knowledge graph.}. ~\label{fig:do}
  \end{minipage}
\end{marginfigure}

Every device and application should identify the associated events and write them to a shared event ledger. For example, a life logging application would find out when you are having a meal and write it to the ledger along with the associated metadata, including location, time stamp, and relevant experiential information \cite{Xie2008}, such as the name of the dish and an image if available.
Figure \ref{fig:do} illustrates how we find the causal relationships from the event ledger and add them to the knowledge graph.

\subsection{Event Miner}
We can use an event mining framework as described in \cite{Jalali2016InteractiveStreams} to find associations between events. We have an event algebra to specify the relationship between different events, and we can use  data driven operators to find the significant event associations from the event ledger. These event associations form the hypotheses for the do-operator. An example hypothesis is $Bike\ \omega_{[2,4]}\ Work$, which means that a bike event is followed by a work event within 2 to 4 hours.

\subsection{DO-operator}
The do-operator \cite{Pearl2018TheEffect} takes us from observing associations to interventions. It can be most easily understood by considering the example of A/B testing in software systems. We present two different versions of the same web page to users with just one difference, such as the position of the submit button to find if the position has any effect on the final outcome. 
\newline
Similarly, in an observational dataset such as the event ledger we need to create fair datasets while controlling for the confounding factors \cite{Li2015FromMining} and test for the significance of the hypothesis across different datasets.

\section{Exploit: Personal Health navigation}
Richard Thaler has described in his Nobel prize winning research how our short term impulses work against the long term goals in economic decision making\cite{Thaler2009Nudge:Happiness}. The same ideology could be applied to health care. Large parts of our daily habits and preferences are driven by our short term satisfaction and conveniences, instead of our long term health goals. We can find these relationships between events using the do-operators and find out how different influencing factors affect our immediate behavior. Different applications can use these relationships to "nudge"\cite{Thaler2009Nudge:Happiness} the user towards their personal health goals.
\newline

\subsection{Ideal Guidance Mechanism}
Every individual has some preferred way of consuming information about different topics that changes how they are influenced by the information. 
A health navigator application can identify such behavioral preferences from the knowledge graph if we have a system monitoring the relevant aspects of the user behavior. For example, we can run an experiment using the do-operator to determine if the user's food choices are being influenced by the pictures they view and like. The navigator can use insights like these to provide the most effective guidance, which, in this instance, could be an instagram page promoting healthy food as opposed to a food blog.

\subsection{Switching Between Mediums}
Any application approved by the user can access the knowledge graph. The knowledge graph can also be used by different devices to provide continuous adaptive navigation towards an individual's health goals, such as when changing mediums (leaving the car system, to a mobile phone, to a wearable). For instance, if an individual skips their morning run, then parts of their daily commute could be replaced with walking to provide the required exercise goal they had set. The knowledge graph should initially be encoded with rules derived from expert knowledge. These can be used to provide a cold-start generic recommendation while we are building the user's preference model. 

\section{Challenges}
There are many research and implementation challenges in building a framework as described. We have built an event miner framework capable of finding common patterns in the data. The next steps are to design and implement the Do-operator to create fair datasets\cite{Li2015FromMining} with respect to the control variables and to find causal relationships between the events.
We also need an event taxonomy that can be used by different devices and applications. 
There are also significant technical and privacy issues to consider while designing the shared event ledger.
\balance{}

\bibliographystyle{SIGCHI-Reference-Format}
\bibliography{Mendeley_Event_Mining}

\end{document}